\documentclass[sigconf]{acmart}

\AtBeginDocument{%
  \providecommand\BibTeX{{%
    \normalfont B\kern-0.5em{\scshape i\kern-0.25em b}\kern-0.8em\TeX}}}

\setcopyright{acmcopyright}
\copyrightyear{2021}
\acmYear{2021}

\acmConference[FIRE 2021]{Forum for Information Retrieval Evaluation}{December 13-17, 2021}{India}
\acmBooktitle{FIRE 2021: Forum for Information Retrieval Evaluation, December 13-17, 2021, India}
\acmISBN{978-1-4503-XXXX-X/18/06}

\begin{document}

\title{Attention based end to end Speech Recognition for Voice Search in Hindi and English}


\author{Raviraj Joshi}
\affiliation{%
  \institution{Flipkart}
  \city{Bengaluru}
  \country{India}}
\email{raviraj.j@flipkart.com}

\author{Venkateshan Kannan}
\affiliation{%
  \institution{Flipkart}
  \city{Bengaluru}
  \country{India}}
\email{venkateshan.k@flipkart.com}

\renewcommand{\shortauthors}{Joshi et al.}

\begin{abstract}
  We describe here our work with automatic speech recognition (ASR) in the context of voice search functionality on the Flipkart e-Commerce platform. Starting with the deep learning architecture of Listen-Attend-Spell (LAS), we build upon and expand the model design and attention mechanisms to incorporate innovative approaches including multi-objective training, multi-pass training, and external rescoring using language models and phoneme based losses. We report a relative WER improvement of 15.7\% on top of state-of-the-art LAS models using these modifications. Overall, we report an improvement of 36.9\% over the phoneme-CTC system on the Flipkart Voice Search dataset. The paper also provides an overview of different components that can be tuned in a LAS based system.

  
 
\end{abstract}




\maketitle

\section{Introduction}
 Voice search on mobile or smart devices has gained a lot of attention in the recent past and the rapid progress in the domain of speech recognition is one of the primary reasons for that. Tech giants have invested heavily in voice-based assistants such as the Google Assistant, Apple Siri, Amazon Alexa, and Microsoft Cortana \cite{hoy2018alexa}. Voice communication is also a more natural form of interaction and engagement as compared to a keyboard-based interface and requires minimal familiarity with the written form of the language.

 Further, in India, keyboard-based interactions have added problems.  The native languages use non-Latin scripts and moreover, the informal conversations tend to have a multi-lingual character to them. The voice interface is significantly more important for e-commerce platforms like Flipkart as it eases the product discovery process, something that has a direct impact on the revenue. The voice search in the local language can help people not comfortable in English to confidently search and order products online. Our work is concerned with Voice search in Hindi and English language primarily describing our Automatic Speech Recognition (ASR) component of the Voice search pipeline. 
 
 An ASR system performs transcription of the spoken language. It can be seen as a function mapping spoken audio to a sequence of words. While it is always desirable to directly perform this conversion from speech to words the amount of training data was a limiting factor to account for this extensive mapping. Therefore instead of modeling words as output, much smaller units called phonemes were used to learn shared representation across words. Phonemes are elementary linguistic units of sound that distinguish one word from another in a language. The model trained to convert speech (acoustic features) to phonemes is termed the acoustic model. A separate pronunciation model was used to map the sequence of phonemes to words. A language model is further employed to select the most likely sequence of words. As the amount of training data has increased end-to-end models have been used to directly convert speech to characters(grapheme), words, or sub-words.    
 
 Traditionally ASR systems were build using a combination of Hidden Markov Models(HMM) and Gaussian Mixture Models(GMM) \cite{bourlard2012connectionist}. With the popularity of Deep Neural Networks (DNNs) these systems transitioned to a hybrid DNN-HMM approach \cite{mohamed2011acoustic}. The recent trend is towards building an all neural approach to ASR \cite{prabhavalkar2017comparison}. These end-to-end systems for ASR fuse individual acoustic,  pronunciation, and language models into a single neural network model. These models directly map speech signals to grapheme (ie, the basic unit of a written language) or word or sub-word sequences. The approaches used in these models include connectionist temporal classification (CTC), recurrent neural network (RNN) transducer, and attention-based encoder-decoder \cite{graves2013speech, graves2012sequence, chan2015listen}. 
 
 In this work, we focus on end-to-end models based on attention. We explore the Listen-Attend-Spell (LAS) model for the task of ASR in the e-commerce voice search domain \cite{chan2015listen}. LAS is attention-based encoder-decoder architecture. The acoustic encoder in this model is termed as Listener and the decoder is termed as Speller. The speller directly emits output symbols in the form of a character or word sequence (depending on the set of subwords that it is trained on). We explore a series of modifications on top of the basic LAS model. These modifications are concerned with the training process aimed to build more robust models for a noisy environment. We employ different attention mechanisms, emphasize the importance of dropout and layer-norm, evaluate multi-pass and multi-objective training, and use external re-scoring. Although results have been reported on LSTM based LAS, the modifications can also be applied to Transformer based models \cite{dong2018speech,karita2019comparative,synnaeve2019end,gulati2020conformer}.       

\section{Related Work}
In this section, we will review literature related to ASR and Indian languages. There has been limited research in the area of Hindi speech recognition due to its low resource nature. Initial works in isolated Hindi speech recognition used HMM-based systems trained using HTK toolkit on very limited datasets \cite{kumar2011hindi,kumar2012hindi,choudhary2013automatic}. The HMM-based systems were also build for continuous speech recognition in Hindi \cite{kumar2014continuous,kumar2004large}. 

More recent works have been biased towards building a single multi-lingual system employing end-to-end speech recognition models. These aim to exploit the usage of multiple related languages to overcome resource constraints. A single multi-lingual LAS based model was trained on 10 Indian languages in \cite{toshniwal2018multilingual}. The resources used in this work were proprietary and consisted of around 3 lakh utterances for each language. The resources for Telugu, Tamil, and Gujarati languages were introduced in INTERSPEECH 2018, Low Resource Speech Recognition Challenge for Indian languages \cite{srivastava2018interspeech}. The multi-lingual system submission to this shared task included GMM-HMM, Time Delay Neural Networks (TDNN), TDNN-LSTM, and RNN-CTC based approaches \cite{fathima2018tdnn,vydana2018exploration,billa2018isi}. Finally, transfer learning-based approaches from resource-rich to low-resource language using RNN-T based system was studied in \cite{joshi2020transfer}. 

Our work is specifically concerned with Hindi and English language and does not use any of the publicly available resources. The English text was transliterated to Hindi using our inhouse transliteration APIs. Hence, the output text is in the Devnagari script for both Hindi and English utterances. We built upon the recent advancements in English ASR \cite{prabhavalkar2017comparison} systems and apply them to the voice search task for Hindi and English. 

\begin{figure}[h]
  \centering
  \includegraphics[width=\linewidth]{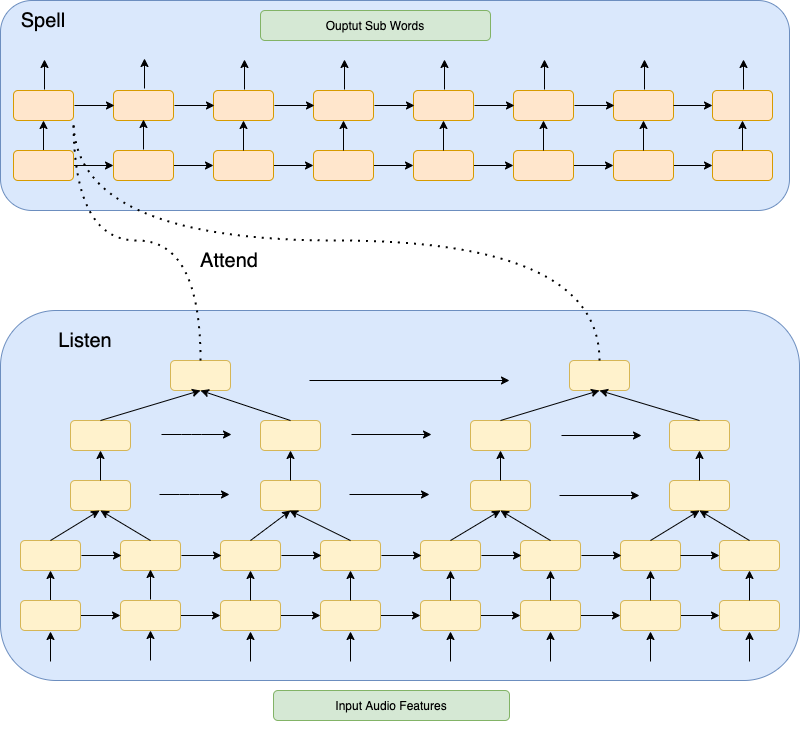}
  \caption{LAS Architecture}
  \label{fig:las_arch}
  \Description{Basic LAS architecture used in our work}
\end{figure}

\section{Listen-Attend-Spell}
\label{sec:las}
LAS is based on the encoder-decoder family of sequence-to-sequence architectures \cite{sutskever2014sequence}. These seq-2-seq models accept variable-length sequence as input and output is also a variable-length sequence. Some of the classic applications of seq2seq models are machine translation, document summarization, speech to text, and text to speech \cite{jean2014using, nallapati2016abstractive, ren2019fastspeech}. The encoder and decoder are typically a stack of recurrent neural networks (RNN/LSTM) although recently Transformers have replaced RNN-type units in most of the tasks. The addition of attention mechanism in the encoder-decoder framework has significantly improved the performance of these models \cite{bahdanau2014neural}. The encoder encodes the input sequence into a sequence of latent representation. The decoder learns to attend to the relevant segment of the entire encoder representation at each output time step by means of an attention mechanism. In other words, the attention mechanism creates a context vector at each output time step (representing the attended information of the encoder sequence) to be utilized by the decoder for generating the next output token. 

LAS is also an encoder-decoder architecture with attention for speech-to-text conversion. More formally it converts a sequence of input $X = \{x_1, x_2, x_3, ..., x_n\}$ into a sequence of output tokens $Y = \{y_1, y_2, y_3, .., y_n\}$. The input is a sequence of spectral features - typically log mel-spectrogram of audio calculated over moving overlapping windows of the audio utterance. The output is a sequence of graphemes (or other sub-word units) in the corresponding script (Devanagari in this case). Two special symbols, the start of sequence <sos> and end of sequence <eos> are added to the start and end of the output sequence respectively. The output at each time step $y_i$ is modeled as a conditional distribution over previous tokens $y_{<i}$ and the input signal $X$. The probability of the entire sequence is modeled as :
\begin{equation}
    P(y|x) = \prod_{i} P( y_i | X,y_{<i})
\end{equation}

The encoder or listener is a bi-directional LSTM (Long Short Term Memory) that encodes input spectra into hidden representation $h$.
\begin{equation}
    h = Listen(X)
\end{equation}
As the input sequence can be considerably long, the listener uses a pyramidal structure to reduce the overall computational complexity. This pyramidal structure involves a multi-layer Bi-LSTM that, at any layer besides the initial one, concatenates two consecutive time-steps of the previous layer to half the number of timesteps. There are four Bi-LSTM layers and the output is compressed after the second, third, and fourth layer to achieve a compression of 8 times. The following equations describe the normal Bi-LSTM layer and pyramidal Bi-LSTM denoted as pBi-LSTM. These demonstrate the computation of output for $i$-th encoder timestep and $j$-th layer.
\begin{align}
    h_{i}^{j} &= BiLSTM(h_{i-1}^{j}, h_{i}^{j-1}) \\
    h_{i}^{j} &= pBiLSTM(h_{i-1}^{j}, [h_{2i}^{j-1}, h_{2i + 1}^{j-1}])
\end{align}

The attend and spell module consumes this hidden representation and previous output token to generate a probability distribution over a set of output tokens for each output step.
\begin{equation}
    Y = AttendAndSpell(h, Y)
\end{equation}
The speller is a 2-layer uni-directional LSTM and the attend function uses Bahdanau Attention mechanism \cite{bahdanau2014neural}.  It uses the speller(decoder) hidden state $s_i$ and context vector $c_i$ to generate distribution over next token. The attention context is computed as
\begin{align}
\label{equ:bah1}
    r_{i,u} = Energy(s_i, h_u) &= v^{T} \tanh{W_1s_i + W_2h_u + b}  \\
    \label{equ:bah2}
    \alpha_{i,u} &= \frac{e^{r_{i,u}}}{\sum_{u} e^{r_{i,u}}} \\
    \label{equ:bah3}
    c_i &= \sum_{u} \alpha_{i,u} h_u 
\end{align}

where $u$ runs over the encoder steps; $W_1, W2, v$ and $b$ are the learnable parameters of the attention mechanism. The attention mechanism computes the relevance of each encoder time-step using the current decoder state. The context vector is a expected value of the encoder representations with associated probability distribution computed by the attention mechanism. The context vector is used by speller to compute the distribution over next token $y_i$ as follows
\begin{align}
    s_i &= \text{LSTM}(s_{i-1}, y_{i-1}, c_{i - 1})\\
    P(y_i|X,y_{<i}) &= \text{DenseSoftmax}(s_i, c_i)
\end{align}

An abstract representation of the basic LAS architecture used in this work is shown in Fig. \ref{fig:las_arch}. This is slightly different from the original LAS description in terms of the number of layers and pyramidal compression. We use 5 Bi-LSTM encoder layers with pyramidal compression after every two layers. This gives us 4x compression in time dimension as opposed to 8x used in the original work. This was done because the voice search queries are comparatively shorter in length (a mean of 5.5 sec) and 4x compression was sufficient.

Some structural and optimization improvements over the base LAS were suggested in \cite{chiu2018state} resulting in state-of-the-art results on the Google Voice Search task. We have incorporated these changes in the base LAS model and considered it as the new baseline. This set of changes along with their corresponding hyperparameters are described below. Some of these modifications did not work for us directly we have also highlighted our design choices to tackle them. 
\begin{itemize}
    \item \textbf{Wordpiece models}: The base LAS uses graphemes or character units as the output token at each step. The output units can also be entire words or phonemes. Systems using words as output units suffer from out of vocabulary (OOV) problems while the usage of phonemes requires additional grapheme to phoneme conversion systems. The middle ground is to use sub-word units that are longer than character units and do not suffer from the OOV problem. The sub-words units allow us to model a strong decoder having lower perplexity as compared to character units and also requires fewer timesteps thus reducing memory and time overhead. Given the superior performance,  all the models described in this work output sub-word units. We use the Google sentence piece library to train a unigram sub-word model and use it for output tokenization \cite{kudo2018sentencepiece}. The dataset used to train this model is the text corresponding to the Voice Search speech to text parallel corpus. The sub-word vocabulary size used is 5k.
    \item \textbf{Muti-headed attention}: The original implementation of the attention mechanism employed a single head to create a single context vector. Multi-headed attention uses multiple attention heads to create multiple context vectors. These context vectors are concatenated to create the final context vector. The importance of scaled dot product multi-headed attention was first shown in \cite{vaswani2017attention} for the task of neural machine translation. We have used multi-headed bahdanau attention instead of scaled dot product attention as it provided better performance for our ASR task. In general multi-headed attention allows the model to focus on different regions of the audio in a better way as compared to single-headed attention. The equations \ref{equ:bah2} and \ref{equ:bah3} are modified as follows to incorporate independent attention heads indexed by $j$.
    \begin{align}
        \alpha_{i,u}^{j} &= \frac{e^{r_{i,u}^{j}}}{\sum_{u} e^{r_{i,u}^{j}}} \\
        c_{i}^{j} &= \sum_{u} \alpha_{i,u}^{j} Z^{j}(h_u) \\
        c_{i} &= [c_{i}^{0}, c_{i}^{1}, c_{i}^{2}, .., c_{i}^{K}]
    \end{align}
    Each attention head has independent learnable parameters; the size of each head is 128 units. The function $Z$ in the above equation is a dense layer having weights and bias as parameters.
    \item \textbf{Scheduled Sampling}:
    The decoder of seq2seq models uses the previously predicted token and context vector to predict the next token. During training instead of using the last predicted token, the correct last ground-truth token is passed to the model. This method is known as teacher forcing and it is known to aid model training during the initial epochs. However, this also results in a disparity between training and the inference mode. During inference time we have to rely on the last predicted token. Again to bridge this gap scheduled sampling was introduced in \cite{bengio2015scheduled} for training RNN based sequence to sequence model. In this approach, at each decoder step, with a sampling probability $p$ the last predicted token will be passed as input instead of the ground truth token. 
    
    The exact process as described in \cite{chiu2018state} used teacher forcing for the first 1 million training steps and post this scheduled sampling is enabled with a probability of 0.4. However, we found this abrupt introduction of scheduled sampling problematic in our case leading to convergence issues. We, therefore, introduce scheduled sampling gradually. 
    
    We have used a maximum scheduled sampling rate of $p=0.3$ and it is enabled after 20 epochs. The sampling rate is linearly increased with a rate of 0.02 per epoch from 0 to 0.3 starting from the 20th epoch. The scheduled sampling parameters - the rate of increase and the starting epoch -  were empirically determined for our dataset ensuring that the model is sufficiently trained before enabling scheduled sampling. 
    \item \textbf{Minimum Word Error Rate Training}:
    The cross-entropy (CE) loss function is used in the training of LAS. The CE loss at each decoder time step between the predicted token and ground truth token is summed over for the entire sequence. However, there is a disparity between the final metric word error rate (WER) and the loss function used to optimize the network. In order to bridge this gap, a minimum word error rate (MWER) loss function is used along with the CE loss function. The MWER loss minimizes the expected value of word errors across all the possible sequences or hypotheses.
    \begin{align}
        \label{equ:mwer}
        \mathcal{L}_{MWER}(x, y^{*}) &= \mathbb{E}[\mathcal{W}(y, y^{*})] = \sum_{y} p(y|x) * \mathcal{W}(y, y^{*}) \\
        \mathcal{L}_{total} &= \lambda * \mathcal{L}_{CE} + \mathcal{L}_{MWER}
    \end{align}
    The function $\mathcal{W}(y, y^{*})$ gives the number of word errors between hypothesis $y$ and ground truth $y^{*}$.
    The equation \ref{equ:mwer} is intractable as we need to sum over all the possible candidate hypotheses. To solve this problem N-best list approximation is used as described in \cite{povey2005discriminative}. We experiment with $\lambda = \{0.3, 0.1, 0.01\}$.
    \item \textbf{Label Smoothing}:
    Label smoothing is a model regularization technique aimed to make model predictions less confident \cite{szegedy2016rethinking}. The idea is to make the output labels noisy by giving a very small non-zero probability to the non ground truth labels. This is shown to have a regularizing effect thereby reducing overfitting. The new label distribution $q'(k)$ for $K$ classes in terms of original distribution $q(k)$ (one-hot) is given by
    \begin{equation}
        q'(k) = (1 - \epsilon) * q(k) + \frac{\epsilon}{K}
    \end{equation}
    The $\epsilon$ is set to 0.1 in our experiments.
    \item \textbf{Second pass rescoring}:
    Second pass rescoring refers to a shallow fusion of the model with an external language model (LM). Although LAS architecture involves an implicit LM in the form of its contextualized decoder, it is trained on a limited amount of transcripts. We can leverage a large amount of textual unlabelled data to build a stronger LM. The usage of this external LM for rescoring the beams from the models has shown to give reasonable improvements. We train an n-gram language model using KenLM \cite{heafield2011kenlm} for external rescoring.
\end{itemize}

\section{Connectionist Temporal Classification (CTC)}
In this section, we provide a brief introduction to CTC-based models. The CTC (encoder only) model monotonically maps input sequence to output sequence of shorter length. CTC is alignment-free and does not require alignment information between audio and text. Each input frame is mapped to a target character (token). The number of input frames is typically much higher than the target tokens (each input frame would correspond to about 20ms of audio), it uses a special blank token and repeat tokens when the spoken segment corresponding to these tokens spans multiple frames. This leads to multiple possible alignments for the same output sequence. The CTC criterion sums the probabilities over all possible alignments to get the log-likelihood of the target sequence $y$. Specifically,
\begin{equation}
    \log p(y|x_{1:T}) = \sum_{a \in \beta^{-1}(y)} \prod_{t=1}^{t=T} p(a_t|x_t)
\end{equation}
where $\beta^{-1}$ is a mapping operation that removes the blank token and repeating tokens to get the final sequence. The probability $p$ can be computed efficiently using a forward-backward algorithm. To get the frame-wise character distribution a simple LSTM network can be used. Therefore a CTC network can be visualized as a network of stacked LSTM layers trained using the CTC criterion.

\section{Methods}
\label{sec:methods}
This section describes the proposed methodology to further enhance the LAS model.
\subsection{Dropout and Layer normalization}
The dropout and layer norm are standard techniques used in deep learning to reduce overfitting and to improve the training time respectively \cite{ba2016layer, srivastava2014dropout}. However, these basic techniques provide us with considerable improvements over the baseline LAS and hence we discuss them separately. The domain of voice search for e-commerce is quite large. There are millions of products that cannot be completely captured in the training data. Moreover, product terms will suffer from under-representation and over-representation in the dataset. The problem of overfitting is quite prevalent with LAS models in such circumstances. The usage of such regularization techniques, therefore, helps us move towards better generalization on unseen data. We apply a block of layer normalization followed by dropout after each Bi-LSTM layer in the encoder and on the output of the two uni-LSTM layers in the decoder before the dense layer. A dropout rate of 0.3 is found to be optimal.
\subsection{Multi-pass training}
The voice search data used for training is a noisy dataset. The noise can either take the form of a background sound or a secondary speaker. This motivated us to explore training using comparatively clean data sets. 
These datasets will be described in later sections. We, therefore, perform multi-pass training to train the LAS model. In the two-pass approach, the model is first trained using comparatively cleaner datasets followed by the training on target voice search data sets. We show that such a training scheme works comparatively better than voice search only training or mixed training. We also explore three pass training where the model is first trained using clean datasets, followed by a mix of clean and voice search data and final training on only voice search data. In this pre-training approach, we use a fixed number of epochs to perform initial training. This fixed number of epochs is set to 150 and (150 + 50) epochs for the two-pass and three-pass approaches respectively. 
\subsection{Multi-objective training}
The LAS model is trained using the cross-entropy criterion. The advantage of LAS over the CTC-based model is that it explicitly exploits conditional dependence using the decoder architecture. The CTC-based models predict output at each encoder timestep independently. However, CTC-based models enforce monotonic alignment between audio and text which is not explicitly captured in LAS models \cite{kim2017joint}. We explore joint multi-objective training using both CTC and LAS CE criteria. The encoder layers are shared between both the CTC and LAS-based models. For the CTC model, a dense layer is added on top of encoder output followed by softmax which generates the label distribution over sub-words for each encoder time-step. These models are jointly trained and the loss function is defined as 
\begin{equation}
    \mathcal{L}_{total} = \lambda * \mathcal{L}_{CE} + (1 - \lambda) \mathcal{L}_{CTC}
\end{equation}
The value of $\lambda$ is set to 0.8. During decoding, we only use the LAS network. This is also a form of multi-task learning wherein the model is also trained on auxiliary tasks to improve the generalization of the model \cite{ruder2017overview}. In this case, the auxiliary task and parent task are the same except that they utilize different objective functions. Another form of multi-objective training is already described earlier using the MWER objective function.
\subsection{Location Aware Attention}
The Bahdanau attention described in equations \ref{equ:bah1}, \ref{equ:bah2}, and \ref{equ:bah3} can also be termed as content-based attention. The attention vector thus computed is dependent upon the values of key and query vectors. A complementary attention mechanism that takes the location of previous attention into consideration was described in \cite{chorowski2015attention}. This form of attention is more suitable for ASR tasks because of the monotonic alignment between speech and text. The location-aware attention is formulated as follows:
\begin{align}
    \label{equ:loc1}
    f_i &= F * \alpha_{i-1} \\
    \label{equ:loc2}
    r_{i,u} &= Energy(s_i, h_u, f_{i,u}) \\
    &= v^{T} tanh(W_1s_i + W_2h_u + W_3f_{i,u} + b)
\end{align}
The $\alpha_{i-1}$ denotes the previous alignment vector. The vector $\alpha_{i-1}$ is convolved with the matrix $F$ to get features corresponding to each encoder time step. The features are then passed to the energy function along with query vector $s_i$ and key vector $h_u$. The variables $W_1, W_2, W_3, v, b$ are the parameters of the model. This attention mechanism is also used in a multi-headed setup as described earlier.

\subsection{Phoneme-CTC based rescoring}
The output of LAS-based models is not explicitly dependent on the length of the audio. The model relies on the generation of <eos> token to mark the end of the sentence. This might cause the LAS model to generate extra undesirable tokens or miss on some tokens. 
To overcome this problem we use a CTC-based model trained to predict phoneme units to re-score the beams generated by the LAS model. The phoneme-based model is expected to penalize the addition of unwanted characters or deletion of required characters. It consists of 5 layer LSTM architecture (700 units) followed by a dense layer. This is used in conjunction with the external language model for final rescoring. The final score for each beam element is computed using the following equation 
\begin{multline}
    final\_score = \alpha * las\_score + \\
    (1 - \alpha) (\beta * lm\_score + (1 - \beta) * phoneme\_ctc\_score)) 
\end{multline}
The values for $\alpha$ and $\beta$ are determined using grid search on a separate validation set. 

\begin{table*}
  \caption{Word Error Rate(WER) for different model variations}
  \label{tab:wer}
  \begin{tabular}{ccccl}
    \toprule
    Model & Rescoring & clean test & noisy test & \vtop{\hbox{\strut clean test + }\hbox{\strut noisy test}} \\
    \midrule
    Phoneme CTC system & \checkmark & 12.77 & 16.9 & 14.87\\ \hline
    Base LAS (single pass)& & 10.94 & 13.03 & 12.0\\
     & \checkmark & 9.95 & 12.25 & 11.12\\ \hline
    Base LAS (single pass) dropout + layernorm & & 9.49 & 11.56 & 10.54\\
     & \checkmark & 9.12 & 11.16 & 10.15\\ \hline
    LAS + two pass + dropout + layernorm (LASR) & & 8.87 & 10.58 & 9.74 \\
     & \checkmark & 8.63 & 10.37 & 9.51 \\ \hline
    LASR + Location Attention & \checkmark & 8.43 & 10.41 & 9.44\\
    LASR + Joint CTC objective & \checkmark & \textbf{8.42} & 10.5 & 9.47\\
    LAS + three pass + dropout + layernorm & \checkmark & 8.5 & \textbf{10.22} & \textbf{9.37}\\
  \bottomrule
\end{tabular}
\end{table*}

\section{Experiments}
\subsection{Dataset Details}
Two datasets are used in this work viz. general domain speech data and Flipkart voice search app data. Flipkart is a primary e-commerce player in India and severs products in electronics, fashion, beauty, grocery, lifestyle, and other categories. The voice search app data consists of audio samples collected from the Flipkart application manually transcribed by the operations team. It consists of approximately 3 million samples with equal distribution of Hindi and English sentences. The general domain data is a comparatively clean dataset and has around 4 million crowdsourced samples. Taken together we have approximately 4000 hours of VS data and 6500 hours of general domain data. For experiments involving two-pass training, the initial training is done on general data followed by noisy VS app data. In the three-pass approach, general data training is followed by (general data + VS app data) and final only VS app data training. 

The test and validation data consists of 9593 and 10000 samples respectively from the voice search domain. The test data is further divided into two categories depending upon the presence of noise or secondary speaker in a test sample. The first category is termed as 'clean test' and consists of only intelligible primary speaker speech. The second set, termed 'noisy test', consists of a primary speaker and secondary speaker but the secondary speech is not intelligible. The first set consists of 4675 examples and the second has 4918 examples. We do not consider the cases where secondary speech is intelligible.     
\subsection{Preprocessing}
The input features to the model are standard log-Mel spectrogram. The spectrogram is computed using a window size of 20 ms and an overlap of 10 ms. The number of mel filterbank coefficients used is 80. We augment the input features by applying time and frequency masking as described in \cite{park2019specaugment}.  A random mel-frequency range of $[f_o, f_o + f)$ is masked and time range of $[t_o, t_o + f)$ is masked using this process. The parameters used for masking are $f=27$ and $t=100$. Features from three consecutive audio frames are concatenated to give a single encoder timestep feature of size 240.

\begin{figure*}[h]
  \centering
  \includegraphics[scale=0.5]{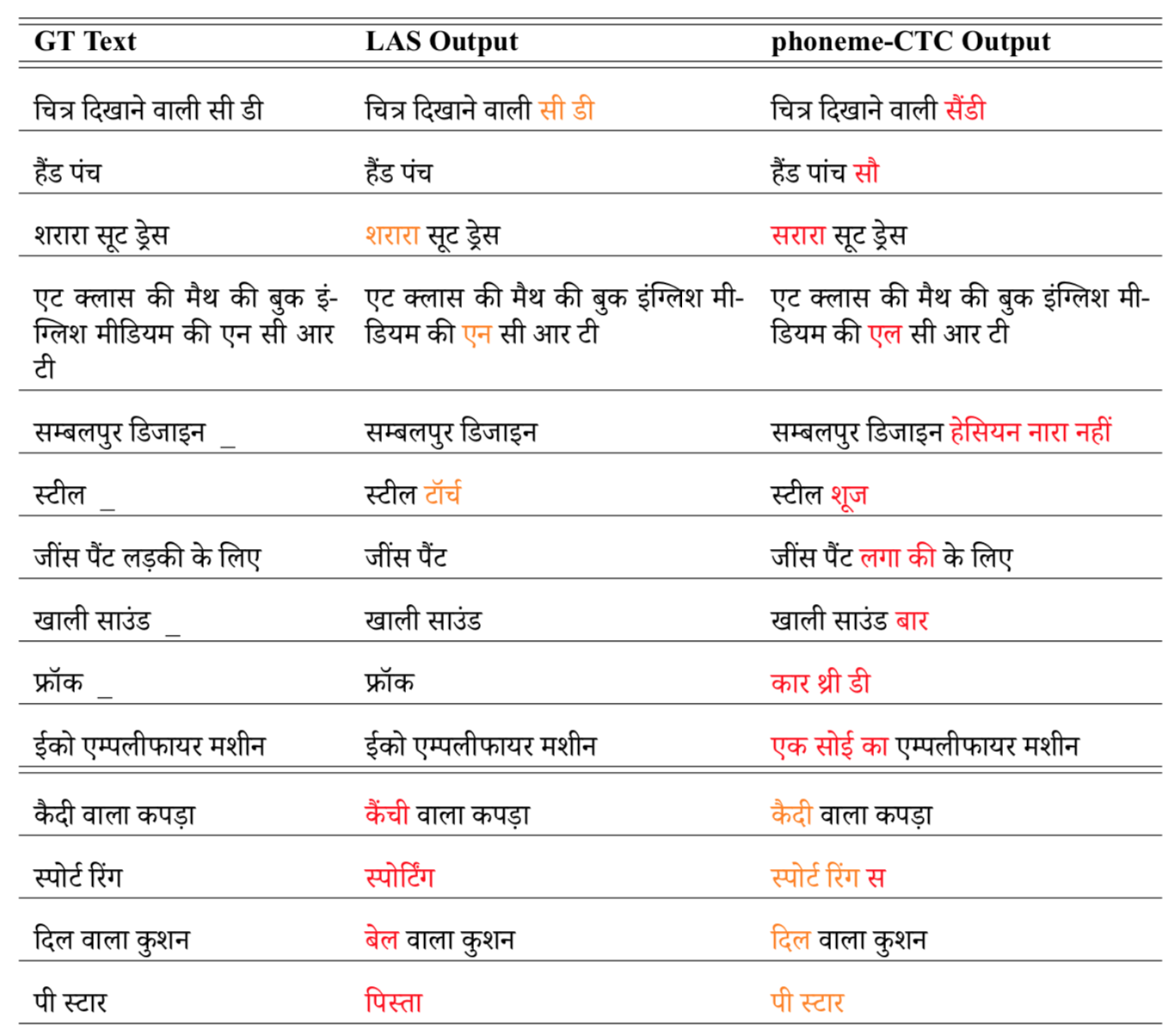}
  \caption{Sample output from the best LAS and phoneme-CTC system. GT Text is the ground truth text. The $\_$ in GT indicates the presence of background noise. The utterances for which the LAS model performs better are placed at the top whereas the utterances transcribed correctly by the phoneme-CTC system are placed at the bottom. The red color indicates incorrect token (modification or addition) and the orange color indicates its correct counterpart. }
  \label{fig:las_output}
\end{figure*}

\subsection{Results and Discussion}
Table \ref{tab:wer} describes the results of the experiments discussed in Section \ref{sec:las} and \ref{sec:methods}. The word error rate (WER) on test data is used to compare these model variations. The pyramidal LAS model along with modifications like scheduled sampling, label smoothing, MHA, and sub-word unit outputs is termed as Base LAS. The baseline system uses the phoneme-CTC model and its WER numbers are reported for comparison. It was made sure that these systems were trained on the same data with the same preprocessing and post-processing in order to have a fair comparison between these systems. The inference was done on GPU for all the systems and had comparable latency requirements. We discuss the relative improvements with the model modifications on the entire test set. During decoding standard beam-search algorithm was used with a beam size of 10.

The Base LAS system reports 25.2\% relative performance improvement over the phoneme-CTC system. We see that the addition of dropout and layer norm to the Base LAS system improves the performance by 8.7\%. The addition of two-pass training utilizing transcription data further improves the results by 6.3\%. The multi-objective training and location-aware attention were independently added on top of this two-pass model. These additions provide a small improvement of around 0.7\%. Finally, the three training approach provides an improvement of 1.5\% as compared to the two-pass approach.

We also see that external rescoring provides an improvement of 7.3\% on the Base LAS system and 2.3\% on the two-pass system. The phoneme and LM-based rescoring contribute equally to these improvements.  

Overall the best configuration reports an improvement of 36.9\% over the phoneme-CTC system and 15.7\% over the Base LAS model. Also, note that relative improvements on clean data are 33.4\% (wrt prod) and 14.57\% (wrt Base LAS) as opposed to 39.5\% and 16.5\% on the noisy data. This shows that LAS-based models and the modifications are more robust to noisy data. Some sample transcriptions from the LAS model and phoneme-CTC system are shown in Fig. \ref{fig:las_output}.
\section{Conclusion}
In conclusion, we demonstrate the superior performance of attention-based speech recognition models on the Flipkart Voice Search Task. We incorporate a series of enhancements on top of the latest Listen-Attend-Spell model in the form of regularization, multi-pass training, multi-objective training, and external rescoring. We show that these simple yet effective structural cum data-related modifications give us a significant boost in performance.

\bibliographystyle{ACM-Reference-Format}
\bibliography{main}










\end{document}